\documentclass[modern]{aastex61}

\newcommand\aastex{AAS\TeX}

\shorttitle{\aastex\ The Massive Protostar in IRAS$\,$18566+0408}
\shortauthors{Hofner et al.}

\begin{document}

\title{High Resolution Observations of the Massive Protostar in IRAS$\,$18566+0408}

\correspondingauthor{Peter Hofner}
\email{phofner@nrao.edu}

\author{P. Hofner}
\altaffiliation{Adjunct Astronomer at the National Radio Astronomy Observatory, 1003 Lopezville Road, Socorro, NM 87801, USA}
\affil{Physics Department, New Mexico Tech, 801 Leroy Place, Socorro, NM 87801, USA}

\author{R. Cesaroni}
\affiliation{INAF, Osservatorio Astrofisico di Arcetri, Largo E. Fermi 5, 50125 Firenze, Italy}

\author{S. Kurtz}
\affiliation{Instituto de Radioastronom\'ia y Astrof\'isica, P.O. Box 3-72 Universidad Nacional Aut\'onoma de M\'exico, Morelia 58090, Mexico}

\author{V. Rosero}
\affil{Physics Department, New Mexico Tech, 801 Leroy Place, Socorro, NM 87801, USA}
\affiliation{National Radio Astronomy Observatory, 1003 Lopezville Road, Socorro, NM 87801, USA}

\author{C. Anderson}
\affil{Physics Department, New Mexico Tech, 801 Leroy Place, Socorro, NM 87801, USA}
\affiliation{National Radio Astronomy Observatory, 1003 Lopezville Road, Socorro, NM 87801, USA}

\author{R. S. Furuya}
\affiliation{Institute of Liberal Arts and Sciences, Tokushima University, 1-1 Minami Jousanjima-machi, Tokushima 770-8502, Japan}

\author{E. D. Araya}
\affiliation{Physics Department, Western Illinois University, 1 University Circle, Macomb, IL 61455, USA}

\author{S. Molinari}
\affiliation{INAF-Istituto di Astrofisica e Planetologia Spaziale, via Fosso del Cavaliere 100, 00133 Roma, Italy}


\begin{abstract}
We report  $3\,$mm continuum, CH$_3$CN(5-4) and $^{13}$CS(2-1) line observations with CARMA, in conjunction with
$6$ and $1.3\,$cm continuum VLA data, and $12$ and $25\,\mu$m broadband data from the Subaru Telescope toward the massive proto-star IRAS$\,18566+0408$.
The VLA data resolve the ionized jet into 4 components aligned in the E-W direction. Radio components
A, C, and D have flat cm SEDs indicative of optically thin emission from ionized gas, and component B has a spectral
index $\alpha = 1.0$, and a decreasing size with frequency $\propto \nu^{-0.5}$.  Emission from the CARMA $3\,$mm continuum, and from the $^{13}$CS(2-1), and CH$_3$CN(5-4)
spectral lines is compact (i.e. $<  6700\,$AU), and peaks near the position of VLA cm source, component B. Analysis of these lines indicates hot, and dense
molecular gas, typical for HMCs. Our Subaru telescope observations detect a single compact source, 
coincident with radio component B, demonstrating that most of the energy in IRAS$\,$18566+0408 originates from
a region of size $< 2400\,$AU. We also present UKIRT near-infrared archival data for  IRAS$\,$18566+0408 which show
extended K-band emission along the jet direction. We detect an E-W velocity shift of about $10\,$km$\,$s$^{-1}$ over the HMC in the CH$_3$CN lines possibly tracing the
interface of the ionized jet with the surrounding core gas. Our data demonstrate the presence of an ionized jet at the base of the molecular outflow, and 
support the hypothesis that massive protostars with O-type luminosity form with a mechanism similar to lower mass stars.

\end{abstract}




\section{Introduction}

The study of massive star formation throughout the Galaxy began with radio continuum surveys (e.g.  Downes \& Rinehart 1966) 
which detected many compact, thermal
sources, later identified as Ultracompact HII (UCHII) regions (e.g. Harris 1973). Observations of these dense, and small ionized regions 
culminated in several
interferometric surveys (e.g. Wood \& Churchwell 1989, RMS Survey by Urquhart et al. 2009, CORNISH Survey by Hoare et al. 2012), where hundreds of UCHIIs were imaged at sub-arcsecond resolution. 
While UCHII regions are manifestations
of recently formed massive stars, because of their bright radio continuum emission, it is thought that their central stars are fully formed,  
and located on, or near the main sequence.

To understand how massive stars form, i.e. how they assemble most of their mass, earlier evolutionary stages must be studied. 
With this goal, a number of surveys were carried out
during the last two decades (e.g. Molinari et al. 1996, Shridharan et al. 2002), selecting massive proto-stellar candidates with 
the criteria of dense molecular cores with large FIR luminosity,
in the absence of strong radio continuum emission. A large number of such objects were catalogued and further studied in a variety 
of observational probes,
such as masers (e.g. Hofner \& Churchwell 1996,  Kurtz, Hofner \& Vargas-Alvarez 2004, Araya et al. 2007a), highly excited molecular 
lines (e.g. Olmi et al. 1996, Araya et al. 2005, Rosero et al. 2013), 
as well as low critical density molecular transitions to search for molecular flows (e.g. Zhang et al. 2001, Beuther et al. 2002). One 
particularly important result of these latter two studies
was the extremely high detection rate of molecular flows toward massive protostars, which indicated that outflows are an essential 
ingredient in the formation of massive stars.

In this paper, we present a multi-wavelength, high angular resolution study of a massive proto-stellar candidate
which we have observed in the $3\,$mm continuum,
CH$_3$CN(5-4) and $^{13}$CS(2-1) spectral
lines with CARMA, and at $12$ and $25\,\mu$m with the Subaru 
Telescope. These data are discussed in conjunction with 6 and 1.3$\,$cm VLA continuum data (Rosero et al. 2016),
as well as archival NIR data.

The high-mass  proto-stellar candidate IRAS$\,$18566+0408 (also known as G37.55+0.20, or Mol83)
is located at a distance of $6.7\,$kpc, and
has  a reported FIR luminosity of $6\times 10^{4}\,$L$_\odot$, 
equivalent to an O8 ZAMS star (Shridharan et al. 2002). The radio continuum emission at 6, 3.6 and $1.3\,$cm was studied by 
Araya et al. (2007b) at an angular scale of one arcsecond and rms noise levels of $20 - 100\mu$Jy/beam.
Considering the large luminosity, the cm continuum emission of about 0.7$\,$mJy is much weaker than what is expected from an
UCHII region, and is best explained by an optically thin thermal jet oriented
in the East-West direction. At $7\,$mm the emission is stronger and shows an elongation approximately perpendicular to the jet, 
which was interpreted by Araya et al. (2007b) as a
circumstellar torus, possibly containing an accretion disk at smaller scales (see their Figure~4).

IRAS$\,$18566+0408 is associated with maser emission from the H$_2$O, OH, CH$_3$OH, H$_2$CO molecules (Araya et al. 
2010, and references therein,  Al-Marzouk et al. 2012). Araya et al. (2010) found that the H$_2$CO maser undergoes periodic flares with a period of about 240 
days, and the flares are correlated with similar
features in the 6.7$\,$GHz CH$_3$OH and 6.035$\,$GHz OH masers (Al-Marzouk et al. 2012). A possible explanation for this behavior is 
maser gain changes due to the infrared radiation of periodic accretion from a binary within a circumbinary
disk (Araya et al. 2010; see van der Walt 2014, Inayoshi et al. 2013 for other possible models).

Zhang et al. (2007) studied IRAS$\,$18566+0408 in several NH$_3$ transitions with the VLA, as well as in the SiO(2-1) and HCN
(1-0) transitions with OVRO. 
They discovered  a well collimated molecular flow centered on a compact $87\,$GHz continuum source, MM-1. The direction of the 
flow is in the SE-NW direction and approximately in the same direction as the CO-flow measured at $10^{\prime\prime}$ resolution 
by Beuther et al. (2002). Furthermore, a strong increase of line-width was observed at the
position of MM1, which Zhang et al. (2007) interpreted either as rotation/infall, or relative motion of unresolved proto-stellar cores.

In the following section we describe our observations, and we present the data in section~3. Section~4 contains 
a discussion of the obtained results. The paper concludes with a brief summary in
Section~5.

\section{Observations}
	
\subsection {CARMA Observation}

Continuum and spectral line observations in the $3\,$mm window toward the IRAS$\,18566+0408$ region were carried out with CARMA (Combined Array for
Research in Millimeter-wave Astronomy\footnote{Support for CARMA construction was derived from the states of 
California, Illinois, and Maryland, the James S. McDonnell Foundation, the Gordon and Betty Moore Foundation, the 
Kenneth T. and Eileen L. Norris Foundation, the University of Chicago, the Associates of the California Institute of 
Technology, and the National Science Foundation. Ongoing CARMA development and operations are supported by the 
National Science Foundation under a cooperative agreement, and by the CARMA partner universities.}).
At the time of observations, CARMA was a 15 element interferometer with nine $6.1\,$m antennas and six $10.4\,$m antennas. For details of the
instrument see the CARMA webpage\footnote{http://mmarray.org/}.

Data were taken in CARMA's B configuration with antenna baselines ranging from $65\,$m to $800\,$m
between January  and February 2008. The field of view of our observations, given by
the half-power beam width of the $10.4\,$m antennas, was 73\arcsec, and our maps have a $\sim$ 1\arcsec\ FWHM 
synthesized beam.
The CARMA correlator recorded signals in three separate bands,
each with an upper and lower sideband. Two bands were configured with a $31\,$MHz bandwidth
($100\,$km$\,$s$^{-1}$)
and 63 channels ($1.6\,$km$\,$s$^{-1}$ per channel),
and were used to observe the $^{13}$CS(2--1) ($\nu_0 = 92.494303\,$GHz)
and CH$_{3}$CN(5--4) ($\nu_0 = 91.971310\,$GHz for K=3) rotational lines in the lower sideband.
One additional band was configured with $62\,$MHz bandwidth across 63 channels, which was also used to observe
the CH$_{3}$CN (5-4) transition
in the lower sideband, with a total velocity coverage of $200\,$km$\,$s$^{-1}$ and a channel
width of $3.2\,$km$\,$s$^{-1}$. The corresponding three bands in the upper sideband at a 
mean frequency of $95.9\,$GHz were used to detect the $3\,$mm continuum emission. Care was taken that the
continuum bands were free of any strong line emission.
 
In each source-calibrator cycle, data were obtained in ten seconds records with 8 minutes spent on 
IRAS$\,$18566+0408 
and 3 minutes on each of two calibrator sources. One of the calibrators (1751+096) was used as the gain and bandpass 
calibrator for both IRAS$\,$18566+0408, and the second calibrator (1827+062) was used as a test source to verify 
the quality of the phase transfer and other calibration steps. Radio pointing was done at the beginning of each track and 
every two hours thereafter. Absolute flux calibration was accomplished using the flux of 1751+096 (1.2 Jy) as determined
from monitoring observations during the same time period as our target observations. Based on the repeatability of the quasar 
fluxes, we estimate that the random error in our source fluxes is $\sim$ 5\%, and the systematic error from the planet 
calibration models is also on that order. We applied a line-length correction to account for changes in the delays of the 
optical fibers as they heat and cool during the day/night cycle. Calibration and imaging were done using the MIRIAD data 
reduction package (Sault et al. 1995). The rms noise in the resulting continuum map was 1$\,$mJy$\,$beam$^{-1}$, and
the rms noise in a channel map of the narrow and wide spectral line data was 12, and 8$\,$mJy$\,$beam$^{-1}$, respectively.

\subsection {VLA Observations}

Observations toward IRAS$\,$18566+0408 were carried out with the Karl G. Jansky
Very Large Array (VLA)\footnote{The National Radio Astronomy Observatory is a facility of the National Science Foundation 
operated under cooperative agreement by Associated Universities, Inc.}  at $1.3\,$cm in the B-configuration on March 20, 2011, and at
$6\,$cm in the A-configuration on July 27, 2011, as part of the survey reported by Rosero et al. (2016).
For  further details on the observations and data reduction we refer the reader to the above paper.
The VLA maps used in this work have 
a synthesized beam of $0\farcs33 \times 0\farcs32$,
position angle PA $=93.4^{\circ}$, rms noise of $4.3\,\mu$Jy beam$^{-1}$  at 6$\,$cm,  and $0\farcs 40 \times 0\farcs30$,
position angle PA $=-31.3^{\circ}$, rms noise is $7.0\,\mu$Jy beam$^{-1}$  at 1.3$\,$cm.\\

\subsection {Subaru Telescope Observations}

IRAS$\,$18566+0408 was observed with the COMICS camera on the $8.2\,$m Subaru Telescope on 2011 August 26, in the
12 and $25\,\micron$ continuum bands. For more details on these observations and data reduction we refer the reader
to Beltr\'an et al. (2014).

\begin{figure}[h!]
\plotone{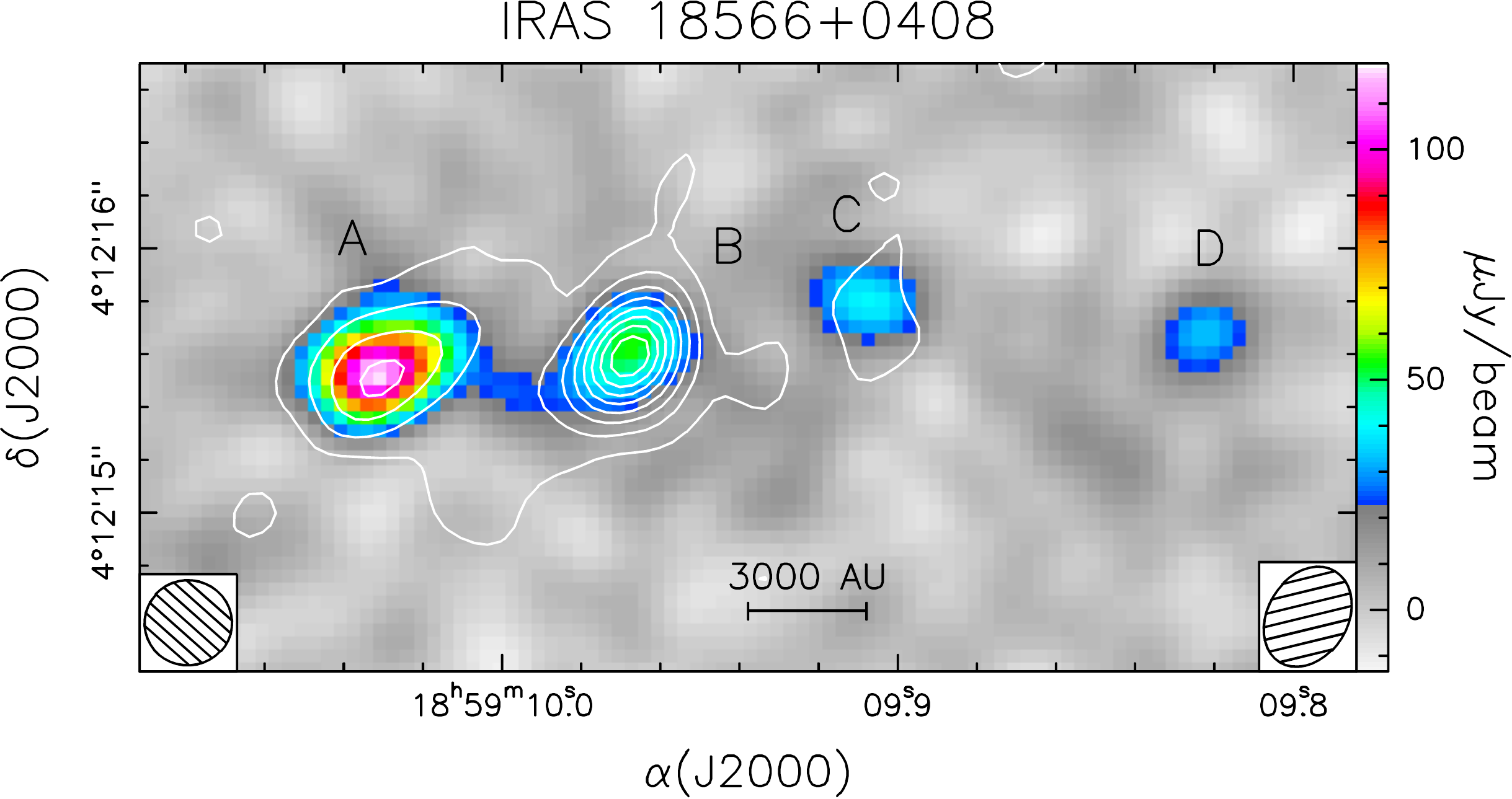}
\caption{VLA continuum data from Rosero et al. (2016). The white contours show the combined $1.3\,$cm map with
contour levels 24 to 192 in steps of $24\,\mu$Jy/beam. The color map shows the combined $6\,$cm map.
The synthesized beam of the $6\,$cm map is shown in the lower left, and that of the $1.3\,$cm map is shown in the lower right 
corner. \label{fig:f1}}
\end{figure}

\section{Results}

\subsection{VLA 6$\,$cm and 1.3$\,$cm Continuum Data}

We show the VLA continuum maps in Fig.~1, with the $1.3\,$cm emission in white 
contours, overlaid on the $6\,$cm map in color. The jet previously observed by Araya et al. (2007b) at angular resolutions of $ 1 - 2^{\prime\prime}$ is now
resolved into 4 and 3 individual components at $6\,$cm and $1.3\,$cm, respectively. These components are labeled
with letters A -- D from East to West (Rosero et al. 2016). At the angular resolution of the VLA data, corresponding to $\approx 2300\,$AU, the sources have a compact core surrounded by  low level 
extended emission. In particular, in both wavelength bands there is an extended structure
connecting the two brighter components A and B, which is fairly narrow at $6\,$cm, but much more extended at $1.3\,$cm.
The length of the East-West structure is about $3\farcs7$ ($25000\,$AU).
Measured peak positions, fluxes and peak intensities for each component are given in Table~4 of Rosero et al. (2016).
If the emission is integrated over the entire
structure we find that the total flux is consistent with the lower angular resolution measurements of Araya et al. 
(2007), i.e. our high resolution maps do not miss any flux. 

\begin{figure}
\plotone{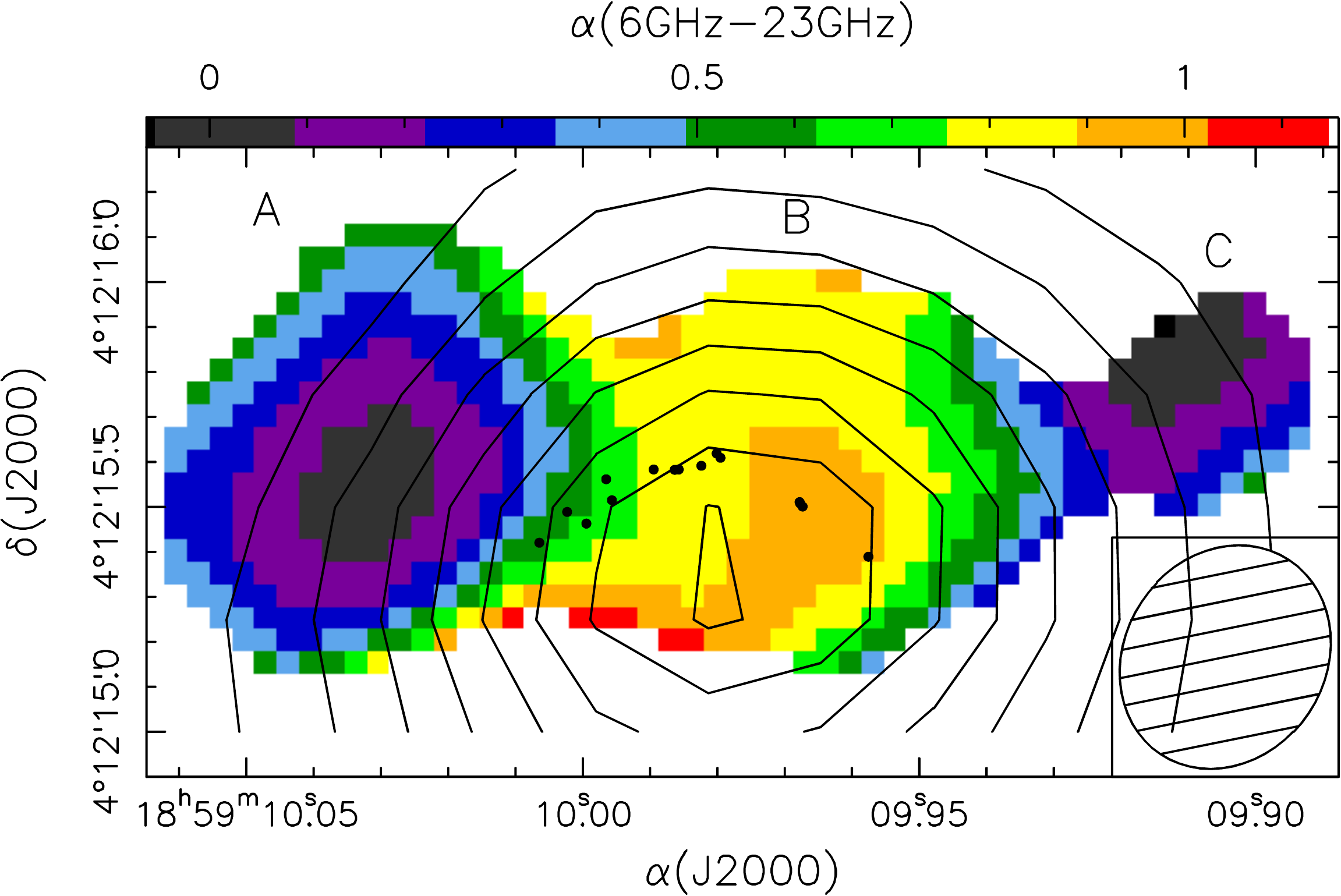}
\caption{ Map of the spectral index of the IRAS$\,18566+0408$ jet, derived pixel-by-pixel for the
combined maps at $6$ and $1.3\,$cm . The contours show the average
CH$_3$CN J = 5--4 K= 0,1 emission. The black dots show the position of the CH$_3$OH $6.7\,$GHz masers
from Araya et al. (2010). The beam for the cm data is shown in the lower right corner. \label{fig:f3}}
\end{figure}

Rosero et al. (2016) show the spectral behavior of the continuum components in their Figure~4.
Components A and C have a fairly flat spectrum, which is indicative of an optically thin
thermal spectrum, and for component D a limit of $\alpha < 0.1$ is reported.
Component B, on the other hand, has a distinctly different behavior. The emission strongly increases at higher frequencies and the fitted
spectral index has a value of 1.0, i.e. in between the values of $-0.1$ for optically thin, and $+2$ for optically thick ionized gas.

A more detailed view of the run of the spectral index $\alpha$ can be seen in Fig.~2. For this map we have smoothed the combined 
$6\,$cm and $1.3\,$cm sub-band maps
to the same resolution and determined $\alpha$ at each pixel. We also overlay on this figure a map of the average
CH$_3$CN J = 5--4 K= 0,1 emission
in contours and the positions of the $6.7\,$GHz CH$_3$OH masers from Araya et al. (2010) as filled circles. From this figure we 
see that the largest spectral index occurs close to the CH$_3$CN maximum which is also near the symmetry axis defined by the CH$_3$OH masers. The 
spectral index drops along
the East-West jet axis, indicating also a drop in optical depth of the ionized gas.

We have fitted 2-D gaussians to the VLA continuum components A - D using the CASA task {\it imfit}.
While components A and B are marginally resolved, components C and D are unresolved.
Component B appears to be slightly extended in the East-West direction, and we find that
the deconvolved source FWHM decreases as as a function of frequency as $\Theta_{major} \propto \nu^\gamma$ where $\gamma = -0.5$. This is shown in
Fig.~3. No variation of size as a function of frequency was detected for component A.

\begin{figure}
\plotone{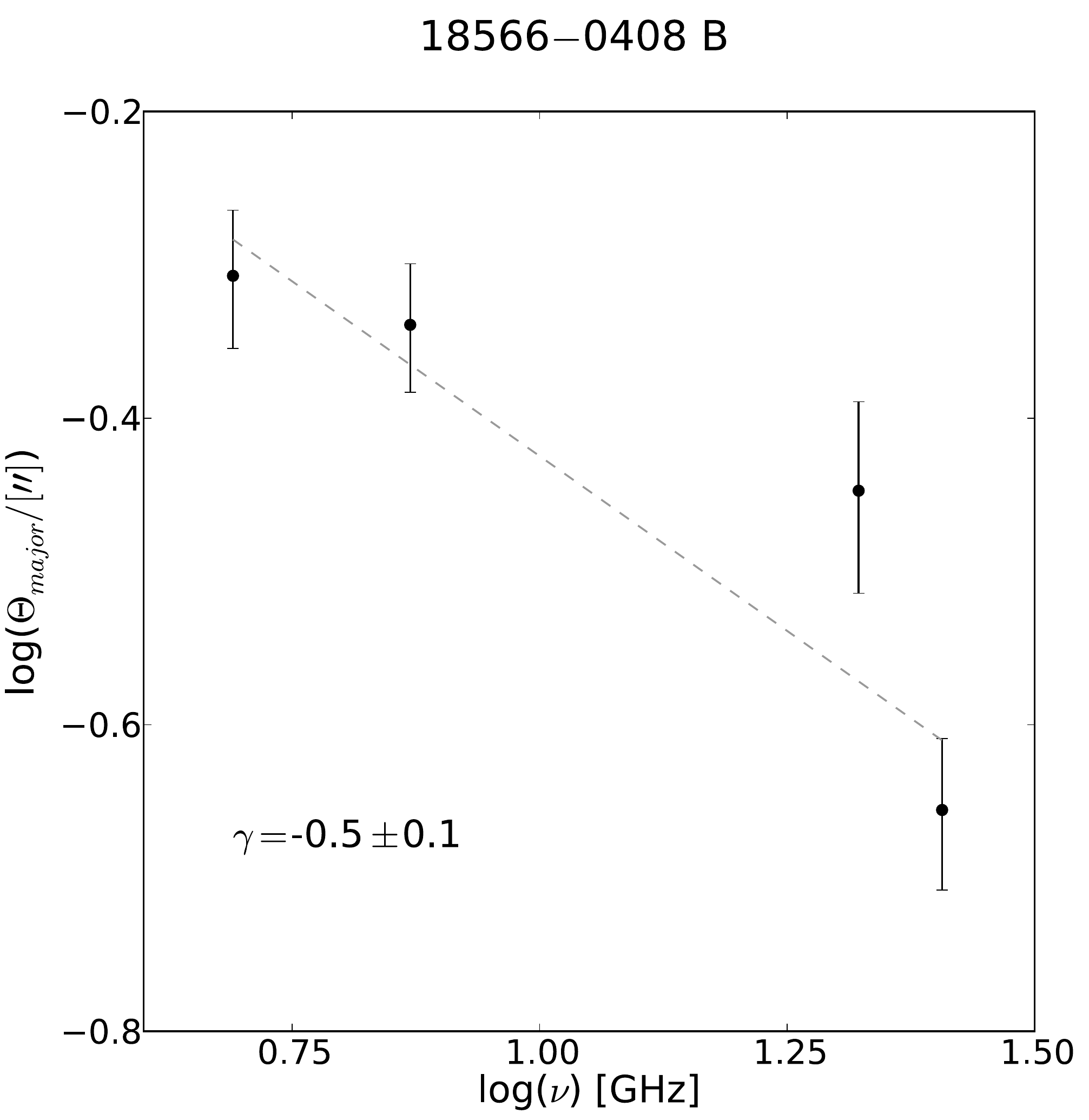}
\caption{ Deconvolved size versus frequency relation for component B. The dashed line shows the powerlaw fit to the data.\label{fig:f4}}
\end{figure}

\subsection{CARMA Data}

\subsubsection{3$\,$mm Continuum}
 
In Figure~4a we show our $3.1\,$mm map in contours overlaid on the $6\,$cm radio jet in color.
We measure a peak intensity of $3.8\,$mJy$\,$beam$^{-1}$, and a total flux density of $7.8\,$mJy. The peak position of the $3.1\,$mm emission is
at R.A. = $18^{h}59^{m}10^{s}.01$, and Decl. = $ +04^{\circ}12^{\prime}15\farcs3$. In Figure~4a we also show the position of the
$7\,$mm peak position from Araya et al. (2007b), which is identical to the position of MM-1 of Zhang et al. (2007).
Considering the positional accuracy of these measurements we conclude that the mm emission peaks are consistent.
The $3.1\,$mm emission in our map is slightly resolved and has a cometary 
shape pointing toward the East, in the direction of the jet.

\begin{figure}
\plotone{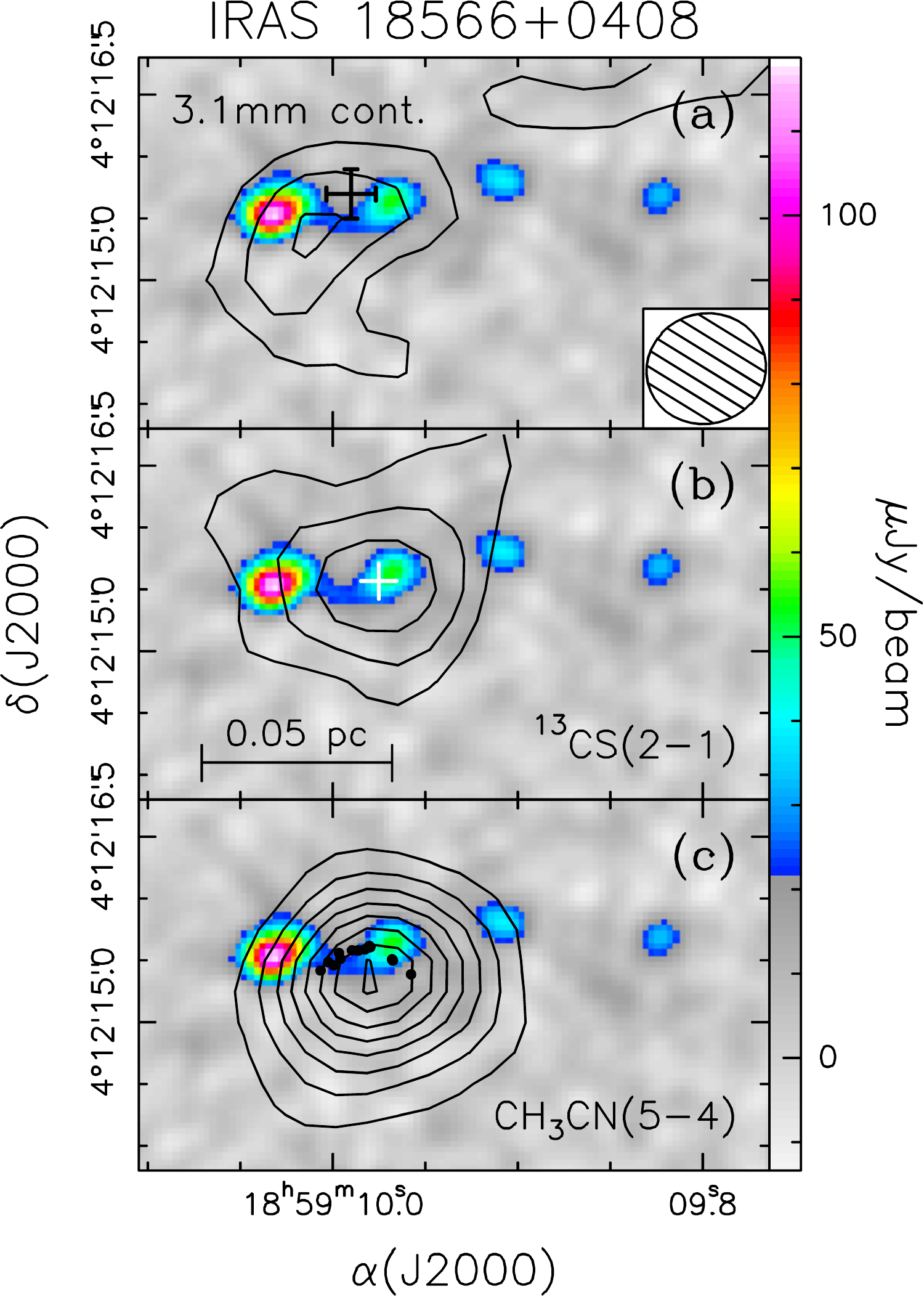}
\caption{ Contour maps of the CARMA data, with the synthesized beam size shown only in the top panel.
 In all images the VLA $6\,$cm map is shown in color in the background.
a) Top panel: Contour map of the $3.1\,$mm emission. Contour levels are from 1.2 to 3.6$\,$mJy$\,$beam$^{-1}$ in steps of $1.2\,$mJy$\,$beam$^{-1}$.
The black cross marks the peak position of the $7\,$mm emission from Araya et al. (2007b).
b) Middle panel: Contour map of the average $^{13}$CS (2--1) line emission.  
Contour levels are from 18 to 42$\,$mJy$\,$beam$^{-1}$ in steps of $12\,$mJy$\,$beam$^{-1}$.
The white cross marks the peak position of the $4.8\,$GHz H$_2$CO maser from Araya et al. (2005b).
c) Bottom panel: The contours show the average CH$_3$CN J = 5--4 K= 0,1 emission. 
Contour levels are from 12 to 96$\,$mJy$\,$beam$^{-1}$ in steps of $12\,$mJy$\,$beam$^{-1}$
The black dots show the position of the $6.7\,$GHz CH$_3$OH masers
from Araya et al. (2010).\label{fig:f5}}
\end{figure}

Zhang et al. (2007) reported $3.4\,$mm observations toward this source with OVRO. Within a $\approx 5^{\prime\prime}$ beam these
authors measure a peak intensity of $18\,$mJy$\,$beam$^{-1}$, and a total flux density of $31\,$mJy integrated over a source size of about $10^{\prime\prime}$.
These data indicate an extended structure which becomes brighter toward the center. Our observations do not have sufficient sensitivity to detect the extended
emission and only trace the central core.
Using the formulas given in Mezger et al. (1990) with $b = 1.9$, solar metallicity, and a temperature of $80\,$K (Zhang et al. 2007), the $3.1\,$mm emission in the central synthesized
beam corresponds to a total mass of about $44\,$M$_\odot$. This is consistent with the estimates based on the NH$_3$ observations of Zhang et al. (2007). 
These authors also reported a secondary peak, MM-2, located to the north-west of MM-1, with a peak brightness of $2.6\,$mJy$\,$beam$^{-1}$ at $87\,$GHz. This
source is not detected in our $95.9\,$GHz observations with a $5\,\sigma$ limit of $5\,$mJy$\,$beam$^{-1}$.

\subsubsection{$^{13}$CS(2--1)}

In Fig.~4b we show the average emission of the $^{13}$CS(2--1) line in contours.
The $3.1\,$mm continuum emission has been subtracted.
The maximum of the line emission is located at
R.A. = $18^{h}59^{m}09^{s}.98$, and Decl. = $ +04^{\circ}12^{\prime}15\farcs5$,
very close to the position of the H$_2$CO  maser (Araya et al. 2005), which is shown as a white cross.
A spectrum taken at the peak pixel is shown in Fig.~5. The line is fitted well by a gaussian with an integrated line flux of
$57\pm7\,$K$\,$km$\,$s$^{-1}$, a velocity of v$_{LSR} = 84.3 \pm 0.3\,$km$\,$s$^{-1}$,
and a line width (FWHM) of $5.2\pm 0.8\,$km$\,$s$^{-1}$.

\begin{figure}
\plotone{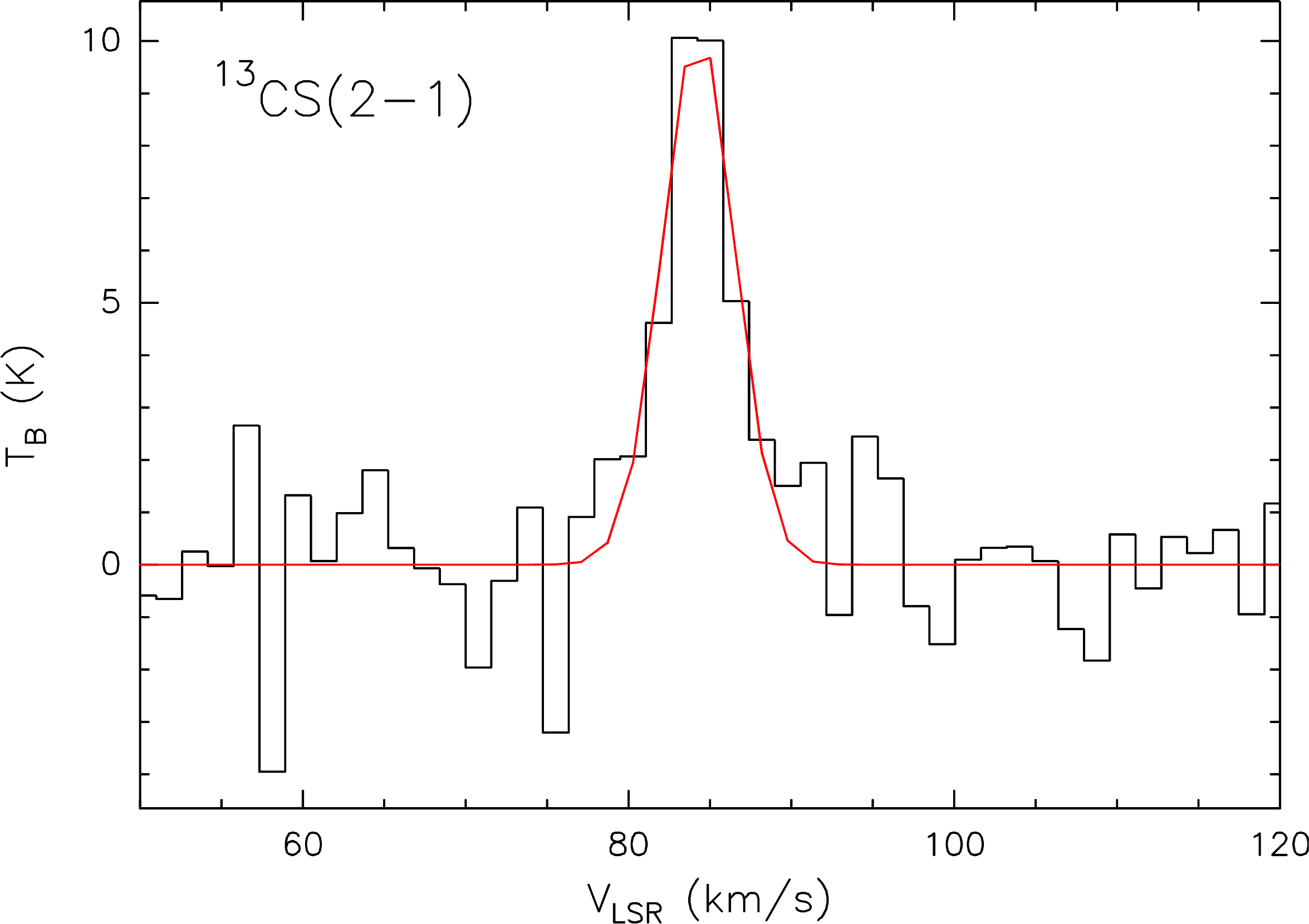}
\caption{CARMA $^{13}$CS(2--1) spectrum toward IRAS$\,18566+0408$.
The spectrum has been extracted at the peak pixel position of
R.A. = $18^{h}59^{m}09^{s}.98$, and Decl. = $ +04^{\circ}12^{\prime}15\farcs5$.
The gaussian fit is shown as the red line.\label{fig:f6}}
\end{figure}

Using the formulas by Mehringer (1995) which assume optically thin emission under LTE conditions, and using an excitation temperature of $80\,$K, the 
integrated line flux at the peak position corresponds to a $^{13}$CS column density of $1.0\times 10^{15}\,$cm$^{-2}$. To estimate the
H$_2$ column density one needs to know the CS abundance, as well as the $^{12}$C/$^{13}$C isotope ratio. While the latter is fairly well known
(we assume a value of 50 here), the CS abundance in hot molecular cores is highly variable due to the active chemistry in these regions.
The models of Nomura \& Millar (2004) predict values between $10^{-7}$ and $10^{-10}$ for the CS abundance. If we adopt a value of
$10^{-8}$, the hydrogen column density and total mass are consistent with what is derived from the $3.1\,$mm emission (see above)
and  implies values of N$_{H_2} = 2.1 \times 10^{24}\,$cm$^{-2}$, and n$_{H_2} = 2.3 \times 10^{7}\,$cm$^{-3}$ within the central $0.03\,$pc.

\subsubsection{CH$_3$CN(5-4)}

In Fig.~4c we show the average of the CH$_3$CN J = 5--4 K = 0,1 emission from the higher spectral resolution
data. The black dots show the position of the CH$_3$OH $6.7\,$GHz masers
from Araya et al. (2010). The CH$_3$CN J = 5--4 K= 0,1 emission has its peak
at a position R.A. = $18^{h}59^{m}09^{s}.98$, and Decl. = $ +04^{\circ}12^{\prime}15\farcs3$.
In Figure~6 we show a spectrum of the low resolution data taken at the peak position.
We have detected all $0 - 4$ K-components of the J = 5--4 transition, as well as marginal emission of the
J = 5--4 K= 0,1 transition of the CH$_3^{13}$CN isotopologue.

\begin{figure}
\plotone{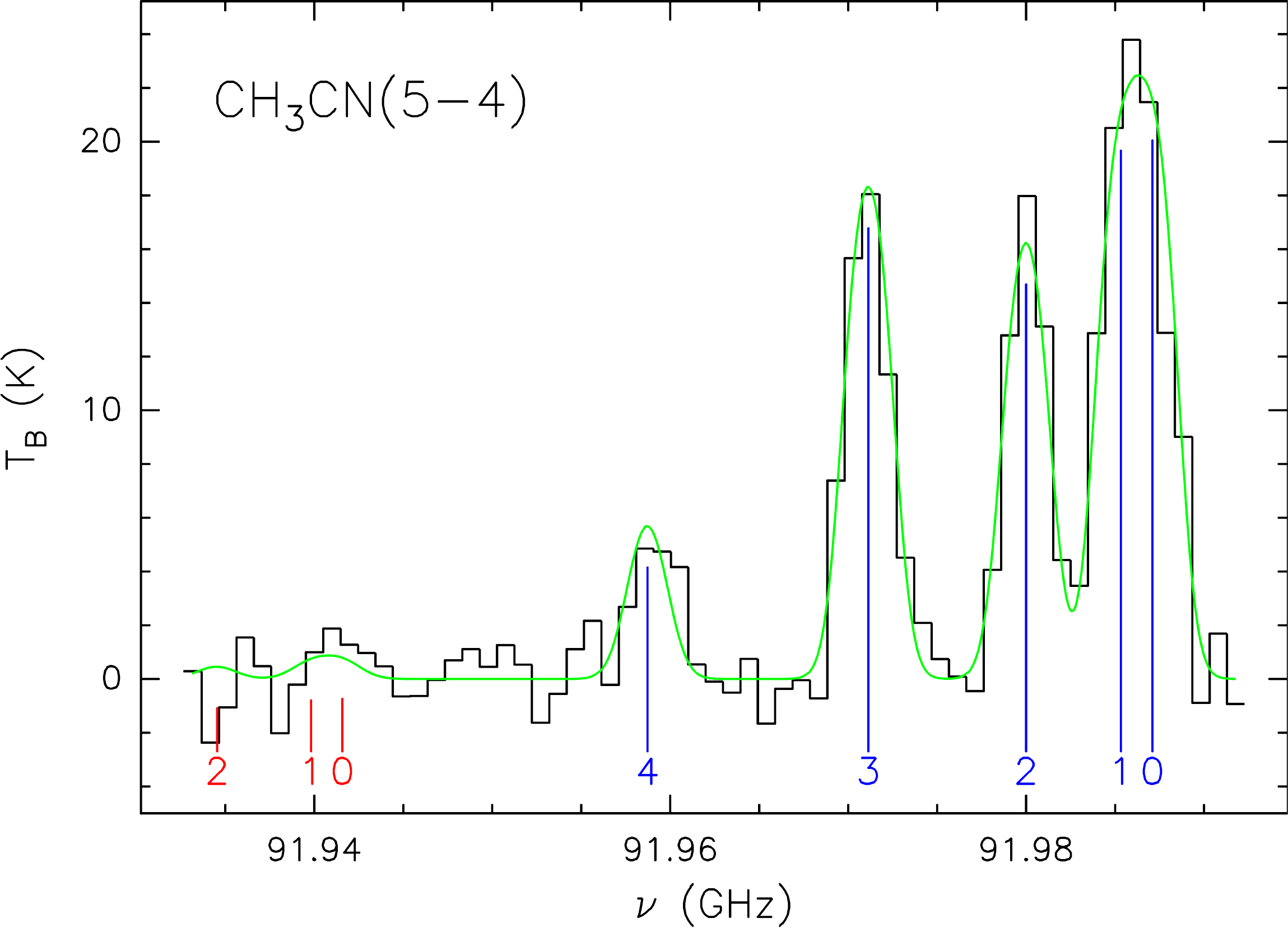}
\caption{ CARMA CH$_3$CN(5--4) spectrum toward IRAS$\,18566+0408$.
The spectrum has been extracted at the peak pixel position of
R.A. = $18^{h}59^{m}09^{s}.98$, and Decl. = $ +04^{\circ}12^{\prime}15\farcs3$.
The green line shows the physical fit described in the text. The frequencies of the K-components of the
CH$_3$CN(5--4) transition are shown in blue vertical lines, and those of the CH$_3^{13}$CN isotopologue  in red line.\label{fig:f7}}
\end{figure}

We fitted the expression $T_B =  T_{rot} \eta (1-e^{-\tau_\nu})$ simultaneously to all detected K-components in the following fashion:
First, a gaussian fit was carried out for the K = 0 -- 4 components with the line separations fixed to their
theoretical values. From this fit we obtained the source velocity v$_{LSR} = 83.9 \pm 0.2\,$km$\,$s$^{-1}$.
Then the expression for the brightness temperature was fit varying the 4 input parameters, beam filling factor $\eta$ ,
line width FWHM, rotational temperature T$_{rot}$, and source averaged CH$_3$CN column density $N_{\rm CH_3CN}$.
The best physical fit is shown in Figure~6 as a green line.
The best fit values are: $\eta = 0.17^{+0.07}_{-0.08}$, $FWHM = 8.3^{+0.9}_{-0.8}$~km\,s$^{-1}$,
$T_{rot} = 170^{+220}_{-60}$~K, $N_{\rm CH_3CN} = (1.1^{+9}_{-0.7})\times10^{17}$~cm$^{-2}$. 
The 1 $\sigma$ errors have been estimated using the method of Lampton et al. (1976). 

Although the CH$_3$CN emission is only marginally resolved we can check for velocity gradients by
fitting the CH$_3$CN(5--4) profiles (from K = 0 to K = 3) pixel by pixel in the map. The result of this procedure is shown in Fig.~7. The typical error on the velocity is $ < 0.4\,$km$\,$s$^{-1}$.
We detect an East-West velocity gradient.

\begin{figure}
\plotone{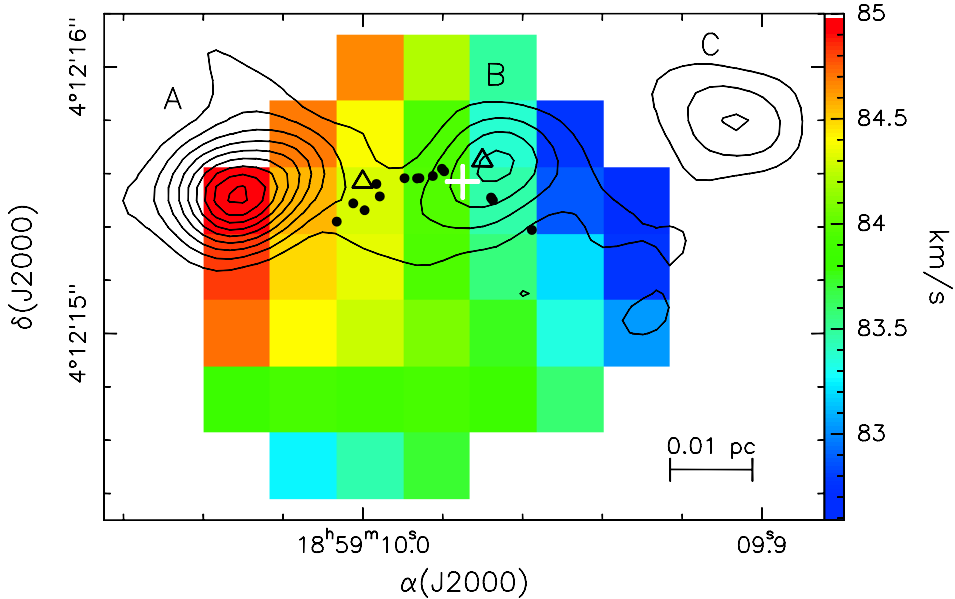}
\caption{ Velocity map of the CH$_3$CN(5--4) K = 0 to K = 3 lines. A spectrum at each pixel was fitted
with fixed linewidth and relative velocities between the K-components for the high spectral resolution data.  An East-West velocity gradient is apparent.
Overlaid are contours of the $6\,$cm emission, while the filled circles are the positions of the $6.7\,$GHz masers
from Araya et al. (2010), the white cross marks the position of the H$_2$CO maser from Araya et al. (2005b), and the
open triangles mark the position of H$_2$O masers (H. Beuther, priv. comm.).\label{fig:f8}}
\end{figure}

An alternative view of the velocity gradient can be obtained by fitting the peak position in each channel map and plotting the 
velocities versus the RA-offsets of the corresponding peak positions.  This is shown in Fig.~8.  A velocity shift
of about $10\,$km$\,$s$^{-1}$ over a distance of $0\farcs4$ ($2700\,$AU) is detected, with redshifted velocities in the East, and
blueshifted velocities in the West.

\begin{figure}
\plotone{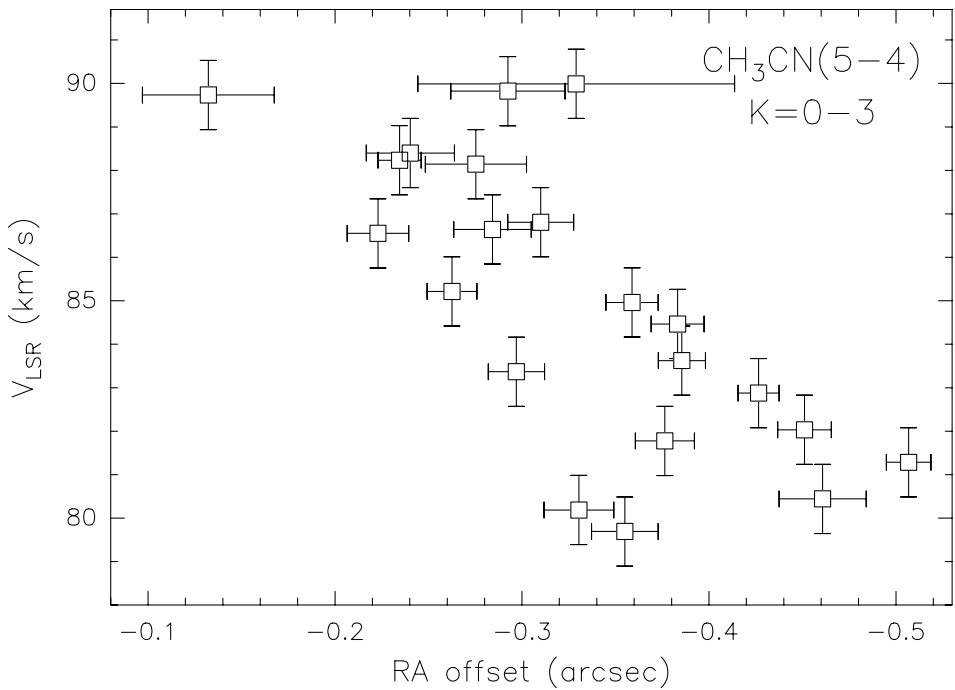}
\caption{Plot of V$_{LSR}$ versus RA offset for the CH$_3$CN(5--4) transition.
The peak position at each velocity was obtained by fitting a 2-D Gaussian to the corresponding channel
map in the high spectral resolution data. Errorbars on the offsets are the formal errors from the 2D-Gaussian fit, while those on the velocities are 
equal to the channel width. A velocity shift of about $10\,$km$\,$s$^{-1}$ over a distance of
$0\farcs4$ ($2700\,$AU) with redshifted velocities in the East, and blueshifted 
velocities in the West is apparent.\label{fig:f9}}
\end{figure}

\subsection{Infrared Data}

In the Subaru images a single point source is detected at both $12$ and $25\,\mu$m, with fluxes of $11\pm 1\,$Jy, and $80\,\pm 8\,$Jy, respectively.
Since this was  the only 
source detected in the field of view of the camera, we registered the position of this source with respect
to the SPITZER/GLIMPSE $8\,\mu$m image. The estimated accuracy of the position of the Mid-IR source is
$1\farcs5$. The source is unresolved with the Subaru diffraction limit of $0\farcs36$ at $12\,\mu$m, which
corresponds to a linear size of $0.01\,$pc or $2400\,$AU at the distance to IRAS$\,$18566+0408 of $6.7\,$kpc.

We also retrieved data from the UKIRT Infrared Deep Sky Survey (UKIDSS) GPS   
for IRAS$\,$18566+0408. 
The UKIDSS project is defined in Lawrence et al. (2007). UKIDSS uses the UKIRT
Wide Field Camera (WFCAM; Casali et al. 2007). The photometric system is
described in Hewett et al. (2006), and the calibration is described in Hodgkin et al. (2009). The pipeline processing and science archive are described in
Irwin et al. (2009) and Hambly et al. (2008).
The UKIDSS images are three magnitudes deeper and have higher angular resolution ($\sim0.4 ^{\prime \prime}$)
compared to 2MASS data. The astrometric accuracy of the UKIDSS data is about 50 mas.

In Fig.~9 we show an overlay of the UKIDSS K-band ($2.2\,\mu$m) image with the VLA $6\,$cm contours. The position of the Mid-IR source is also shown.
In the infrared K-band there is a bright source coincident with the ionized jet. It is not detected at the infrared H or J band, which suggests that it is deeply embedded. 
The source is very extended with respect to the UKIDSS resolution and hence the most likely explanation for the K-band emission
is scattered light.

\begin{figure}
\plotone{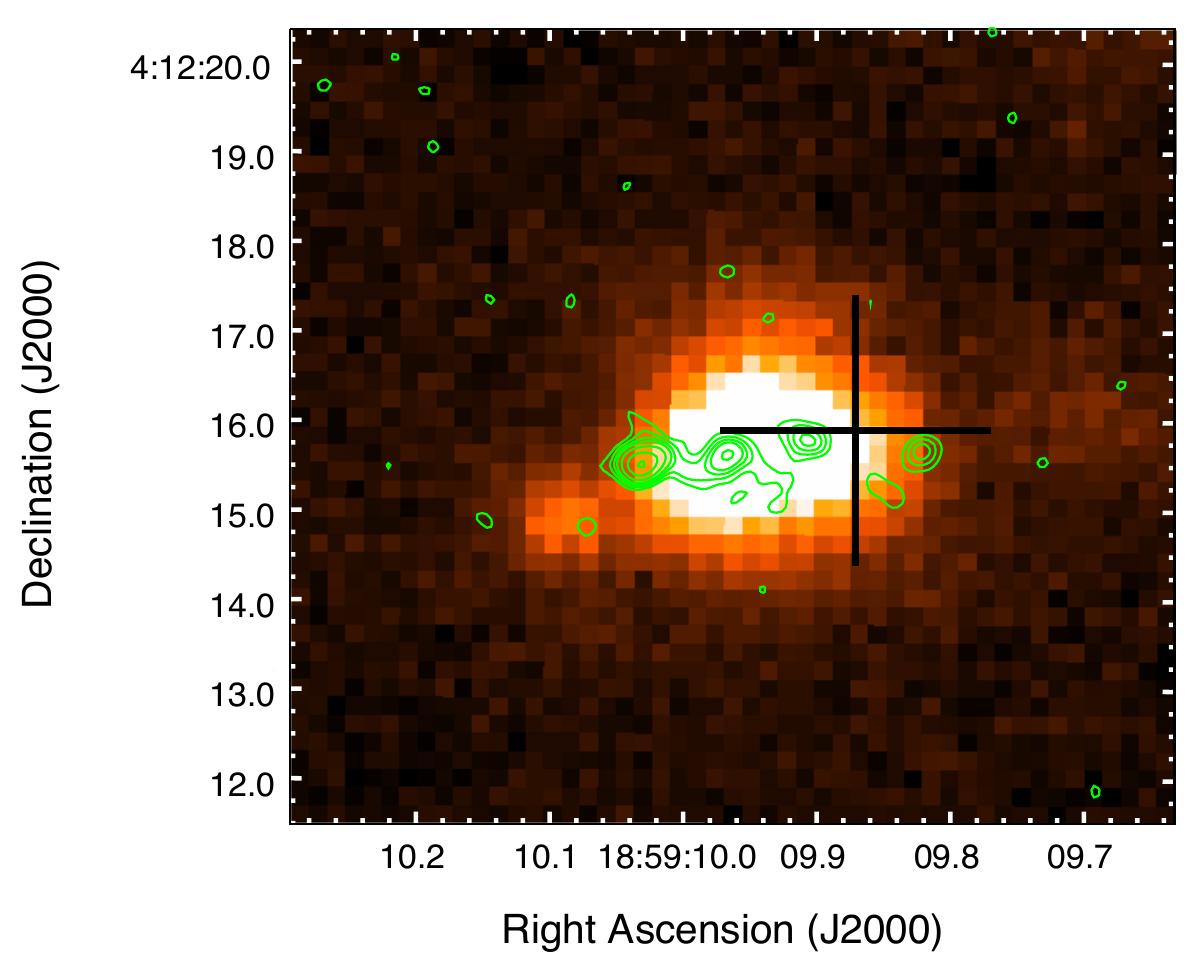}
\caption{Infrared K-band ($2.2\,\micron$) UKIDSS image in color overlaid with the $6\,$cm contours.
Contour levels are
$3, 5, 7, 9, 13, 20, 29 \times 4\,\mu$Jy/beam.  
The black cross marks the position of the $25\,\mu$m emission, and the dimensions of the cross indicate
its approximate astrometric accuracy.\label{fig:f10}}
\end{figure}

We have also revisited the question of the total luminosity of IRAS$\,$18566+0408. For this we collected
all available flux density measurements in the literature and plotted the spectral energy distribution in Fig.~10. 
We obtain a value for the total luminosity of this source of $8\times 10^4\,$L$_\odot$ confirming the results
of Zhang et al. (2007). Inspecting the available maps in the wavelength bands between $24\,\mu$m (SPITZER MIPSGAL) and
$500\,\mu$m (Herschel Hi-GAL) we found that the emission is compact and centered on the position
of the massive protostellar candidate. Furthermore, there are no additional mid-IR
sources within 10$^{\prime\prime}$  from the central source, demonstrating that the compact mid-IR source
dominates the energetics of the region. 

\begin{figure}
\plotone{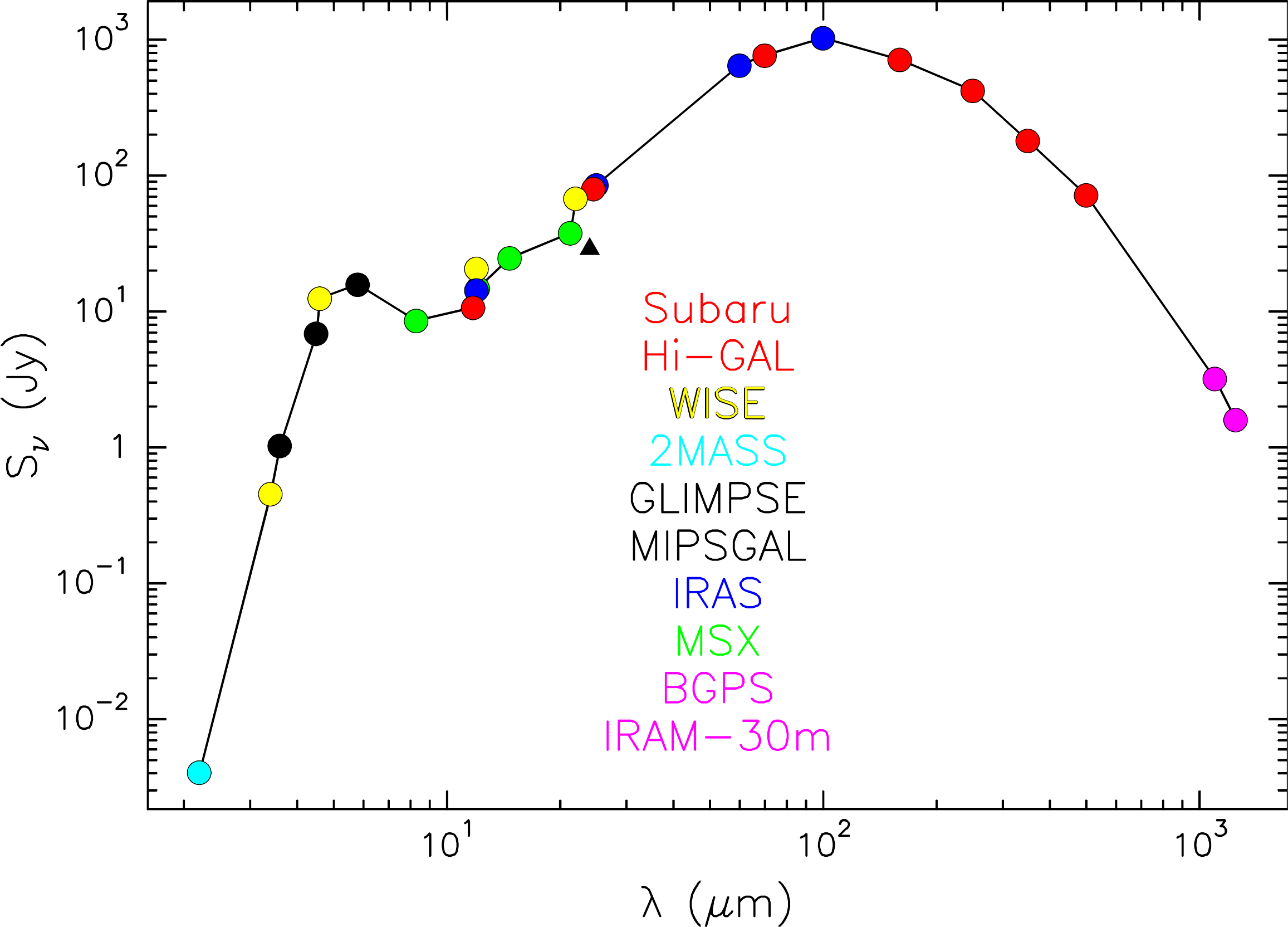}
\caption{Spectral energy distribution toward IRAS$\,18566+0408$. We have collected the
flux densities available in the literature as shown above. The triangle marks a lower limit to the flux density due to saturation.
Integrating under this curve we obtain a
value for the total luminosity of $8\times 10^4\,$L$_\odot$.
\label{fig:f11}}
\end{figure}

 \section{Discussion}

\subsection{The Nature of the cm Continuum Emission} 

Our VLA data show a linear structure oriented East-West of length approximately $0.1\,$pc which consists of weak extended emission
and 4 compact components. One might ask whether these sources could be explained as individual UCHII regions, and although we favor
an ionized jet explanation, we will briefly comment on this hypothesis. First, calculating the required spectral types from the measured fluxes,
assuming optically thin conditions, we obtain ZAMS spectral types of B2 for components A and B, and B3 for components C and D. The total 
luminosity of 4 such stars would be below $2\times 10^4\,$L$_\odot$, whereas the measured luminosity of  IRAS$\,$18566+0408 is
$8\times 10^4\,$L$_\odot$, thus, the regions appear highly under-luminous. However, while absorption of UV photons by dust within 
the ionized gas could in principle explain this discrepancy (e.g. Wood \& Churchwell 1989), in this case one would expect a closer morphological
correspondence in the  mid-IR. As mentioned above, components A, C, and D have flat spectral indices,
indicating optically thin emission, and the data are consistent with emission measures $\approx 10^5\,$pc$\,$cm$^{-6}$ and electron densities of
$\approx 5\times 10^3\,$cm$^{-3}$, as one would expect for small ionized regions around B2--3 type stars. 
For component B the spectral index of 1.0 could be explained with a constant density HII region where the optical depth is approximately unity, i.e.
between optically thick and thin emission. With this assumption we obtain an emission measure of $2.7\times10^9\,$pc$\,$cm$^{-6}$, and 
a lower limit on the electron density of $5.2\times 10^5\,$cm$^{-3}$, similar to values often derived for UCHII regions
(e.g. Kurtz, Churchwell \& Wood 1994).
We have also calculated sizes for the hypothetical HII regions assuming that they are initial Str\"omgren spheres, or ionized gas clouds
in pressure equilibrium with the surrounding gas using the theory of Xie et al. (1996), and the  data of Zhang et al. (2007). In all cases the sources
are predicted to be smaller than our resolution, in agreement with our observations. In summary then, our VLA data do not completely rule out that the
4 continuum sources are individual HII regions. However, due to the association with outflowing molecular gas (Beuther et al. 2002, Zhang 2007),
as well as shocked gas seen as Spitzer green-band excess (Araya et al. 2007b) in approximately the same direction, we consider an ionized
jet a more likely explanation. Our result that component B shows a decreasing size with frequency (Fig.~4) is also an expected feature of varying
electron density in the source (e.g.  Panagia \& Felli 1975, Reynolds 1986), and hence supportive of the ionized jet assumption.
We will hence adopt this hypothesis for the rest of this paper.  

There are several models that can be used to calculate the expected flux from ionized jets. The work of Reynolds (1986) predicts the cm emission from
partly ionized collimated winds without specifying the source of ionization, and the work of Curiel et al. (1987, 1989) 
discusses the case of cm emission from an ionized plasma where a neutral wind shocks molecular matter surrounding the proto-star. Most researchers
have favored the model of Curiel et al. (1989) to explain observed jet sources (e.g. Hofner et al. 2007, Johnston et al. 2013).
Both models depend on many physical parameters which are generally not known, but adopting a number of assumptions we can obtain estimates
for the mass loss ($\dot M$) and momentum rate ($\dot P$) for the jets. Thus, adopting the usual set of assumptions (e.g. Rosero et al. 2014), we can apply equation (19) of Reynolds (1986)
to our radio component B,  to estimate a mass loss rate of ${\dot M} = 1.1\times 10^{-5}\,$M$_\odot\,$yr$^{-1}$, and a momentum rate of
$\dot P = 7.7 \times 10^{-3}\,$M$_\odot\,$yr$^{-1}\,$km$\,$s$^{-1}$. It is interesting to note that Beuther et al. (2002) obtain a similar number for $\dot P$, based on their 
IRAM$\,$30m CO(2--1) maps, which would indicate sufficient force in the jet to drive the larger scale molecular flow.

For the case of shock ionization, Anglada et al. (1992) has shown that for low-mass objects the relation between momentum rate
and cm luminosity follows a powerlaw described by $\dot P = 10^{-2.6}(S_\nu d^2)^{1.1}$ (where $\dot P$ is in units
of $M_\odot\,$yr$^{-1}\,$km$\,$s$^{-1}$, and $S_\nu d^2$ in units of mJy$\,$kpc$^2$),  and Rodriguez et al. (2008) have shown that three high 
luminosity objects with ionized jets also follow this relationship.  For component B at $7.4\,$GHz we have 
a cm luminosity of $2.8\,$mJy$\,$kpc$^2$, which results in a predicted momentum rate of $\dot P = 8.0 \times 10^{-3}\,$M$_\odot\,$yr$^{-1}\,$km$\,$s$^{-1}$, i.e.
virtually identical to the value measured by Beuther et al. (2002), and what is implied by the formula of Reynolds (1986). We hence add an additional data
point to the $\dot P$ vs $S_\nu d^2$ relation, which further strengthens the hypothesis that the ionized jets observed from massive young objects are operating
with the same physical mechanism as those from low mass stars.

\subsection{The Nature of the CH$_3$CN Velocity Gradient}
 
The molecular line data observed with CARMA allow an interesting view on the kinematics in the IRAS$\,$18566+0408 core.
Zhang et al. (2007) detected an increase of FWHM in the NH$_3$ lines from $5.5\,$km$\,$s$^{-1}$ at an angular resolution
of $3^{\prime\prime}$, to $8.7\,$km$\,$s$^{-1}$ at an angular resolution of $1^{\prime\prime}$. They discuss this increase
of linewidth in terms of outflow, infall, rotation or relative motion of multiple objects in the synthesized beam. These authors also mapped a collimated flow
in the SiO(2-1) line centered on the mm peak position which is oriented in the SE--NW direction, with blue-shifted emission extended mostly
toward the NW, and a smaller emission region at red-shifted velocities toward the SE. A similarly oriented flow was also observed by Beuther et al. (2002) in
the CO(2--1) line with the IRAM$\,$30m telescope. 

We have detected a change of velocity in the CH$_3$CN lines along the jet of about $10\,$km$\,$s$^{-1}$ along the East-West direction. This confirms the
broadening of the spectral lines toward smaller scales observed by Zhang et al. (2007). The alignment of the velocity gradient with the ionized
jet strongly suggests that the gas traced by the CH$_3$CN(5--4) line represents outflowing matter at the base of the flow.
We note that the CH$_3$OH and H$_2$O masers are also distributed along an East-West direction, with a velocity gradient consistent with that of CH$_3$CN.  
Since the CH$_3$CN emission is only marginally resolved at an angular scale of $1^{\prime\prime}$, and our fitting technique is essentially sub-resolution, 
the outflowing gas is mostly arising from distances $< 6700\,$AU from the massive protostar.
Our line fitting results indicate a low beam filling factor of $\eta = 0.17^{+0.07}_{-0.08}$, suggesting that most of the emission occurs
on scales of  $\approx 0\farcs5$. We have derived temperatures and column densities of  $T_{rot} = 170^{+220}_{-60}$~K,
and $N_{\rm CH_3CN} = (1.1^{+9}_{-0.7})\times10^{17}$~cm$^{-2}$. We compare these values with the results which were recently
reported by Hern\'andez-Hern\'andez et al. (2014) based on lower resolution SMA CH$_3$CN(12-11) data of this region. 
These authors used a two component model consisting of a lower density cooler region, plus a compact higher density and hotter region to fit their
spatially unresolved spectrum. Due to the relatively large error in our temperature determination, our value for the temperature is consistent with either of these 
components. Our relatively high column density is also consistent with what  Hern\'andez-Hern\'andez et al. (2014) derive from fitting their
compact component if we account for the difference in assumed size for the emitting region. As discussed in Comito et al. (2005), the $\chi^2$ fits
to the spectrum are generally not unique, hence we caution the reader that the size of the emitting region remains uncertain. Higher angular
resolution observations which fully resolve the CH$_3$CN emission will be necessary to clarify this point. 

Adopting then for the remainder of this discussion the physical values derived in our analysis, we can ask what is the likely astrophysical
scenario responsible for the CH$_3$CN emission. The detected E -- W velocity gradient, and the spatial coincidence with the jet, suggests that the CH$_3$CN
emission traces an outflow associated with the ionized jet very near the central protostar.
As discussed above, the momentum rate of the ionized jet appears sufficient to drive the
large scale molecular flow. One possible interpretation of the velocity gradient of the CH$_3$CN lines  is that we have detected the high pressure
neutral gas adjacent to the jet, i.e. the region where the jet imparts momentum and accelerates the molecular gas. Two physical mechanisms for
momentum transport have been discussed in the literature: the formation of a turbulent mixing layer via the Kelvin-Helmholtz instability, or
alternatively via bow shocks (e.g. Chernin et al. 1994, Cant\'o \& Raga 1991). Most jets from young stellar objects are highly supersonic
(e.g. Guzm\'an et al. 2016) and in this case momentum transfer via bow shocks is predicted to dominate (Chernin et al. 1994). Since the 
IRAS$\,$18566+0408 jet appears to lie very close to the plane of the sky, the detected velocity shift along the jet axis of about $10\,$km$\,$s$^{-1}$ indicates
that the 3D space velocities are much higher, and also highly supersonic. We note that while emission in the CH$_3$CN lines has often been 
assigned to a rotating disk perpendicular to the outflowing gas (e.g. Cesaroni et al. 2014), CH$_3$CN emission has been
also reported as tracer of outflows (Leurini et al. 2011 in IRAS$\,$17233-3606, Palau et al. 2017 in IRAS$\,$20126+4104), and of bow shocks (Codella et al. 2009 for L1157-B1).

Our data allow us to compare the thermal pressure of the ionized jet near its base
with the thermal pressure of the surrounding molecular gas, to investigate whether the molecular gas will inhibit the sideways expansion
of the jet and contribute to its collimation.  The ionized gas has a lower limit on the electron density of $5.2\times 10^5\,$cm$^{-3}$ (see above)
and assuming an electron temperature of $10^4\,$K we have P/k$\, \geq 5 \times 10^9\,$K$\,$cm$^{-3}$ (where k is the Boltzmann constant). 
To estimate the pressure in the surrounding molecular gas we need to assume an abundance for CH$_3$CN relative to H$_2$, which can vary between
$10^{-7}$ and $10^{-10}$ in chemically active Hot Core regions (e.g. Nomura \& Millar 2004, Calcutt et al. 2014). Using a common assumption for
the abundance ratio of $10^{-8}$ and a temperature of $170\,$K, the resulting thermal pressure is P/k$\,\approx 7 \times 10^{8}\,$K$\,$cm$^{-3}$,
which is an order of magnitude lower than the limit on the thermal pressure in the jet. It is thus likely that the neutral gas will not inhibit 
sideways expansion of the jet material, and a different collimation mechanism is needed.

\subsection{Comparison with IRAS$\,$16562-3959}

Finally, it is interesting to compare IRAS$\,$18566+0408 with a source which, while located at a much closer distance of $1.7\,$kpc, appears strikingly similar,
namely the jet source IRAS$\,$16562-3959. While many studies of protostars with bolometric luminosities corresponding to
early B-type stars exist, these two sources are rare examples of massive protostars with an O-type luminosity. Both sources have a linear string of radio
sources which are best interpreted as ionized gas due to shocks from fast jets. IRAS$\,$16562-3959 shows radio lobes at distances out to 0.4 pc from 
the central object, and both sources have inner radio lobes at distances of order $10000\,$AU. In the case of IRAS$\,$16562-3959 proper motion of the radio lobes
with approximately $500\,$km$\,$s$^{-1}$ was recently measured by Guzm\'an et al. (2016), proving without doubt their jet origin.
For both sources the radio spectra for the lobes are flat whereas the central object has a rising spectral index of about 1. Also, for both sources, the central source
shows a decreasing size with frequency, which is suggestive of an ionized jet. However, based on RRL data, Guzm\'an et al. (2014) model the radio emission from
the central source in IRAS$\,$16562-3959 as an HCHII region with a slow ionized wind. Clearly, both sources are very rare examples of massive protostars that are still in an accretion
phase during their formation.  One might speculate that IRAS$\,$18566+0408 is perhaps in a somewhat earlier evolutionary state due to its more compact
nature, high level of maser emission, and due to the presence of high pressure molecular gas at the base of the jet.
The presence of jets in both sources strongly supports the hypothesis that the disk accretion model also applies to massive protostars with O-type luminosity.

 \section{Summary}
 
In this paper we presented high angular resolution observation at cm, mm, and mid-IR wavelengths
toward the massive protostellar candidate IRAS$\,$18566+0408. The main results are as follows:
\bigskip


1) Our VLA data in the 6 and $1.3\,$cm wavelength band resolve the ionized jet into 4 components. These
are aligned in the E-W direction consistent with other outflow tracers in this source. Radio components
A, C, and D have flat cm SEDs indicative of optically thin emission from ionized gas, and component B has a spectral
index $\alpha = 1.0$, and shows a decreasing size with frequency with $\gamma = - 0.5$ as expected from 
an ionized jet.

\bigskip

2) We have detected compact (i.e. $<  6700\,$AU) emission in the $3\,$mm continuum, and in the $^{13}$CS(2-1), and CH$_3$CN(5-4)
spectral lines,
which peak near the position of continuum component B. Physical analysis of these lines indicates hot, and dense
molecular gas with values typical for HMCs.

\bigskip

3) Our Subaru telescope observations detect a single compact source at $12$ and $25\,\micron$ which is 
coincident with radio component B. This demonstrates that most of the luminosity in IRAS$\,$18566+0408 originates from
a region of size $< 2400\,$AU.

\bigskip

4) We also present UKIRT near-infrared archival data for  IRAS$\,$18566+0408. The source is only detected at the infrared K-band, and
the extended nature of the emission elongated along the jet direction,  most likely indicates scattered light.

\bigskip

5) We detect an E-W velocity shift of about $10\,$km$\,$s$^{-1}$ over the HMC in the CH$_3$CN lines. It is possible that the
dense and hot molecular gas traces the interface of the ionized jet with the surrounding molecular core gas. If the CH$_3$CN abundance is
assumed to have the typical HMC value of $10^{-8}$, the thermal pressure in the ionized jet exceeds the thermal pressure in the molecular gas,
and the sideways expansion of the ionized gas is not sufficiently inhibited, hence
an additional collimation mechanism may be needed.

\acknowledgements

PH acknowledges support from NSF grant AST-0908901 for this work.
Support for this work was provided by the NSF through the Grote Reber Fellowship Program administered
by Associated Universities, Inc./National Radio Astronomy Observatory (CA and VR).
We thank S. Schnee for help with CARMA data reduction.
Some of the data reported here were obtained as part
of the UKIRT Service Program. The United Kingdom Infrared
Telescope is operated by the Joint Astronomy Centre on behalf
of the UK Particle Physics and Astronomy Research Council.
We thank H. Beuther for providing accurate H$_2$O maser positions.
Further thanks are due to the anonymous referee, whose comments improved
this manuscript.

\end{document}